\documentclass[fp,twocolumn]{jpsj3}
\usepackage{txfonts}

\usepackage{epsfig,bm}
\usepackage[dvips]{color}

\title{Energy and Mass-Number Dependence of Hadron-Nucleus 
Total Reaction Cross Sections}

\author{Akihisa Kohama,$^1$ 
Kei Iida,$^{1,2}$ and Kazuhiro Oyamatsu$^{1,3}$}
\inst{$^1$RIKEN Nishina Center, RIKEN, 
2-1 Hirosawa, Wako-shi, Saitama 351-0198, Japan\\
$^2$Department of Natural Science, Faculty of Science, 
Kochi University, 
Kochi 780-8520, Japan \\
$^3$Department of Human Informatics, Aichi Shukutoku University,
2-9 Katahira, Nagakute, Aichi, 480-1197, Japan
}

\date{\today}
\abst{We thoroughly investigate how proton-nucleus total
reaction cross sections depend on the target mass
number $A$ and the proton incident energy.  In doing so,
we
systematically analyze nuclear reaction data that are sensitive
to nuclear size, namely, proton-nucleus total reaction
cross sections and
differential elastic cross sections, using a phenomenological
black-sphere approximation of nuclei that we are developing.
In this framework, the radius of the black sphere is found to be a
useful length scale that simultaneously accounts for the observed
proton-nucleus total reaction cross section and first diffraction
peak in the proton elastic differential cross section.  
This framework, which is shown here to be applicable to
antiprotons, is expected to be applicable to any
kind of projectile that is strongly attenuated in the nucleus.
On the basis of a cross-section formula
constructed within this framework,
we find that a less familiar $A^{1/6}$ dependence plays 
a crucial role in describing the energy 
dependence of proton-nucleus total reaction cross sections. 
}

\begin{document}
\maketitle

\setcounter{section}{0}
\setcounter{equation}{0}
\section{Introduction} 
\label{intro}

     The total reaction cross section ($\sigma_R$) of nuclei is 
one of the most fundamental observables in nuclear physics, which 
helps us to know nuclear radii and even nuclear density 
distributions.  The role of $\sigma_R$ in deducing the nuclear
density distributions is complementary to that of the 
differential cross section of elastic scattering, and both of them are 
governed by diffraction phenomena~\cite{Bethe}.  On the other hand, 
nuclear masses and radii characterize the bulk properties of nuclei.  
In fact, the saturation of the binding energy and density deduced from 
systematic data for the masses and charge radii of stable nuclei reflects 
the behavior of the equation of state of nearly symmetric nuclear matter 
near the saturation density~\cite{oyaii,iioyaa}.  

As is well known, the nuclear radii and density distributions 
are deduced from
electron and proton elastic scattering off nuclei
~\cite{Alk:PR,Cha:AP,Igo:RMP,Bat:ANP}.
To deduce the matter density distributions and radii,
during the past four decades there have been many efforts of
studying proton elastic scattering cross sections,
which are
based on various scattering theories 
incorporating empirical nucleon-nucleon scattering amplitudes, 
such as the Glauber approximation~\cite{Igo:RMP,Alk:PR} 
and nonrelativistic and relativistic optical potential methods 
~\cite{Ray:PR,Coop:PRC47,Clark:PRC67,sakagu:prc57,zeni}.  

We here choose a different approach: We start from
a naive but firm framework, namely, 
the Fraunhofer diffraction, 
which is expected to set qualitative standards for nuclear size.
Unexpectedly, this framework turns out to be quantitatively
very sound, which will be explained below.

     For the purpose of deducing nuclear size from proton-nucleus elastic 
scattering and $\sigma_R$, we proposed a model in which a nucleus 
is viewed as a ``black'' (i.e., strongly absorptive to incident protons) 
sphere of radius ``$a$", which is called a black-sphere (BS) 
approximation of nuclei~\cite{Bethe,BS1,BS2}.  
This BS radius, $a$, plays a central role 
in this framework.
We determine $a$ by fitting the 
angle of the first elastic diffraction peak calculated 
for proton diffraction 
by a circular black disk of radius $a$ to the measured value. For 
incident protons of energy above 800 MeV, 
it was found that $\pi a^2$ agrees
with the measured $\sigma_R$ within error bars.  It can thus be 
regarded as a ``reaction radius" inside which the reaction with incident 
protons occurs.

     Within the BS framework, we developed a formula for 
$\sigma_R$ of proton-nucleus reactions as a function of the mass
number ($A$) and the neutron excess of the target nucleus and proton 
incident energy $T_p$ in a way free from any adjustable $T_p$-dependent 
parameter~\cite{BS3}.  We deduce the dependence of $\sigma_R$ on $T_p$ 
from a simple argument involving the nuclear ``optical" depth for 
absorption of projectiles.  
We call the formula the BS cross-section formula
~\cite{kurotama}.
The only scale included in the formula is set by the BS radius $a$, 
which is determined in the same way as described above~\cite{BS2}.  
For stable nuclei, this formula reproduces the empirical $T_p$ dependence 
of $\sigma_R$ at $T_p=100$--1000 MeV remarkably well.  In this 
formula, the $T_p$ dependence of $a$ is determined 
by that of proton-nucleon 
total cross sections, while the target mass-number dependence of $a$ is 
sensitive to the surface thickness of the target nucleus.  This formula 
can be easily extended to nucleus-nucleus reactions and is shown to well 
reproduce the empirical data for energies above 
100 MeV/nucleon~\cite{BS3,kio08}. 

     Due to its suitability for systematic calculations, 
the present formula 
is incorporated into the Particle and Heavy Ion Transport code System 
(PHITS)~\cite{iwase,niita,hybkuro}.  
In the code, the formula is used for systematic 
evaluations of $\sigma_R$, which in turn determine how 
often the incident particles collide with nuclei in a material.  The 
application area of this code is very broad, which ranges from 
the fields of accelerator technology, particle therapy, 
and space radiation 
to many other fields that are related to particle and heavy-ion 
transport phenomena. 

     In this paper, 
we revisit a complicated $A$ and $T_p$
dependence of the proton-nucleus total reaction cross sections.
In doing so, 
we will put emphasis on the fact that the BS radius  
is the length scale that simultaneously accounts for 
the observed $\sigma_R$ of proton-nucleus reaction 
and diffraction peak in the proton elastic differential 
cross section. 
After summarizing the successive works on our 
systematic analyses based on the BS approximation of nuclei, we examine 
the $T_p$ and $A$ dependence of $\sigma_R$ carefully.  
A part of the 
results have been already reported in refs.~\cite{BS1,BS2,BS3,kio08}.  

     This paper is organized as follows: 
In Sec.~\ref{bsm}, we overview our  
BS approximation of nuclei.  
In Sec.~\ref{how}, we summarize how data look 
like from the viewpoint of the BS approximation.  
In Sec.~\ref{bsxsec}, we briefly review the BS cross-section formula, 
which was developed in ref.~\cite{BS3}, and analytically examine its 
$A$ dependence.  
Detailed derivation of the formula 
can be found in Sec.\ 3 of ref.~\cite{kurosupple}.
We extend this framework to such probes as 
antiprotons in Sec.~\ref{other-probes}.  Finally we give a 
summary in Sec.~\ref{summary}.

     In collecting the empirical data, we have made access to 
Experimental Nuclear Reaction Data File (EXFOR) \cite{exfor}.  As for the 
criterion to adopt the data for $\sigma_R$, we have 
accepted the data which are to within 15 \% from the systematic 
behavior of various data sets.  We use units in which 
$\hbar=c=1$.

\setcounter{equation}{0}
\section{Black-Sphere (BS) Approximation} 
\label{bsm}

     In this section, we introduce the BS approximation of 
nuclei~\cite{BS1,BS2}, which can be regarded as a ``contemporary" 
BS model as compared with the original one~\cite{Bethe} 
(hereafter referred to as ``classical"). 
We regard the former as contemporary, 
because it is based on 
the quantitative 
reproducibility of available $\sigma_R$ data, 
while the classical one 
aims at qualitatively describing a global behavior 
of the elastic diffraction patterns.  
The formal definition of the contemporary BS approximation 
will be given below via Eq. (\ref{form-a}). 

For convenience, 
we restrict ourselves to the case 
of proton projectiles, 
but
the concept can be easily extended to 
such hadronic probes as neutrons, antiprotons, pions, and 
kaons.  The case of antiprotons will be described in 
Sec.~\ref{other-probes}.  Possible extension of this framework to 
proton inelastic scattering has been discussed recently~\cite{inela},
but will not be discussed here.

We here emphasize that our contemporary BS model is {\it not} 
the eikonal approximation with the rectangular density distribution 
although the connection with it can be clearly shown 
as in Sec.~2 of ref.~\cite{kurosupple}.
Since, for stable nuclei, 
the accuracy of $\sigma_R \simeq \sigma_{\rm BS}$ 
of proton-nucleus 
($A \ge 3$)
and nucleus-nucleus reactions 
($A_P,A_T \lesssim 50$),
where $A_{P(T)}$ is the mass number of a projectile (target), 
has been confirmed within a few \%~\cite{BS2,BS3},  
the indication by Alkhazov {\it et al}.~\cite{alkha:ijmp,novikov} 
that the results of our BS model are not accurate enough 
particularly for light nuclei 
is not appropriate in the context of $\sigma_R$.

Blair {\it et al}. developed the celebrated 
``sharp cutoff" model 
for low energy alpha-particle elastic scattering off nuclei 
several decades ago,~\cite{Blair,FBl}
which is a strong absorption model 
that can be obtained from wave optics 
by cutting off the interaction range or 
the partial-wave (impact parameter) window. 
This model
reproduces a global behavior of 
the alpha-nucleus scattering fairly well.~\cite{Blair,FBl}
In fact, 
the BS approximation is similar 
in concept and structure to the ``sharp cutoff" model,
but how to relate between them is not obvious 
partly because the geometrical size of alpha particles is
treated differently and partly because the definition of the
``sharp cutoff" radius is based on the behavior of partial
waves rather than the nuclear density distribution.

\subsection{Applicability} 

     We begin by regarding a target nucleus for proton elastic scattering 
as a black sphere of radius $a$.  This picture holds when the target 
nucleus is strongly absorptive to the incident proton and hence acts like 
a black sphere.  It is important to notice that the interaction between 
the incident proton and the target nucleus is strong but not infinitely 
strong; otherwise the incident proton could be sensitive to an 
exponentially low density region, and hence any place would 
be black. 

     For incident kinetic energy $T_p$ above $\sim800$ MeV, 
the optical potential for this reaction is in fact strongly absorptive.  
It can be essentially viewed as a superposition of the nucleon-nucleon 
scattering amplitudes.  Since the imaginary part of the amplitude is 
dominant over the real part in this energy range, the BS picture is 
applicable to a first approximation.

     Another requirement for the BS picture is that the proton 
wave length is considerably shorter than the nuclear size.  For proton 
incident energies higher than about 800 MeV, both requirements are 
basically satisfied.  This approximation was originally used by Placzek 
and Bethe \cite{Bethe} in describing the elastic scattering of fast 
neutrons.

     Since one can regard the proton beam as a plane wave of momentum 
$p_{\rm Lab}$ in the laboratory frame,
\begin{equation}
   p_{\rm Lab} = \sqrt{(T_p+m_p)^2-m_p^2}
\end{equation}
with the proton mass, $m_p$, the BS approximation can be described in 
terms of wave optics.   This picture reduces to a diffraction of the 
wave by a circular black disk of radius $a$ if the corresponding wave 
optics is close to the limit of geometrical optics, i.e.,
\begin{equation} 
   \frac{a}{\lambda_{\rm Lab}} \gg 1, 
\end{equation} 
where $\lambda_{\rm Lab}$ $= 2\pi/p_{\rm Lab}$ is the wave length. 
We will consider in the next section the range of $T_p$ 
in which $a/\lambda_{\rm Lab} \gg 1$ is satisfied.  According to 
Babinet's principle, this diffraction is in turn equivalent to 
the Fraunhofer diffraction by a hole of the same shape 
as the disk embedded in a screen \cite{Landau}.  

     The scattering amplitude for this diffraction in the 
center-of-mass (c.m.) frame of the proton and the nucleus reads
\begin{equation}
    f({\bf q})=i p a J_1(qa)/q,
\label{dif}
\end{equation}
where ${\bf q}$ is the momentum transfer, ${\bf p}$ is the proton 
momentum in the c.m.\ frame, and $J_n(x)$ is the $n$-th order Bessel 
function.  With this amplitude, we obtain the differential cross 
section of proton-nucleus elastic scattering as 
\begin{equation}
   \frac{d\sigma}{d\Omega} = |f({\bf q})|^2.  
\end{equation}
The relation of the BS approximation to the conventional 
scattering theory can be found 
in Sec.\ 2 of ref.~\cite{kurosupple}.

     We note that the BS picture is fairly successful 
in describing the elastic scattering of low energy $\alpha$ 
particles \cite{Blair,FBl,Bat:ANP}.  It was also used for analyses of 
the scattering of intermediate-energy pions and 
low-energy antiprotons \cite{Bat:ANP}.

\subsection{How to Determine ``$a$"} 
\label{howto-a}

     The scale ``$a$" is the only undetermined parameter 
in the scattering amplitude of Eq.~(\ref{dif}).  We determine it 
using the empirical differential cross sections 
of proton-nucleus elastic scattering.

     The c.m.\ scattering angle for proton elastic scattering is 
generally given by 
\begin{equation}
   \theta_{\rm c.m.} = 2\sin^{-1}(q/2p). 
\end{equation}
For the proton diffraction by a circular black disk of radius $a$, 
we can calculate the value of $\theta_{\rm c.m.}$ at the first 
peak as a function of $a$.  (Here we define the zeroth peak as that 
whose angle corresponds to $\theta_{\rm c.m.}=0$.)  We determine 
$a$ in such a way that this value of $\theta_{\rm c.m.}$
agrees with the first peak angle for the measured diffraction in
proton-nucleus elastic scattering, $\theta_M$.  The radius, $a$, 
and the angle, $\theta_M$, are then related by
\begin{equation}
   2 p a \sin(\theta_M/2) = 5.1356 \cdots.
    \label{a}
\end{equation}
This is obtained by requiring that the derivative of the 
cross section with respect to the scattering angle be
zero.  To be explicit, we write
\begin{equation}
   a = \frac{5.1356 \cdots}{2 p \sin(\theta_M/2)}.
    \label{form-a}
\end{equation}
We call this the BS radius formula.  As was discussed 
analytically in the eikonal approximation~\cite{ADL}, 
the oscillation period in the diffraction pattern is 
determined by the nuclear radius, which is closely related with 
the concept underlying Eq.\ (\ref{form-a}).

The determination of $a$ from the first peak angle, rather
than the first dip angle, is the key to the success of the
present quantitatively sound approach.

\subsection{Definition of $\sigma_{\rm BS}$ and $r_{\rm BS}$}

     Within the present BS approximation, we calculate the 
proton-nucleus total reaction cross section, $\sigma_R$,  
from $a$.  This approximation regards it as the 
geometrical cross section, 
\begin{equation}
  \sigma_{\rm BS}\equiv\pi a^2.
\label{rrf}
\end{equation}
Here we assume that the incident protons are point particles, 
leading to vanishing contribution from the proton size to 
$\sigma_{\rm BS}$~\cite{BS2}.  This is reasonable 
because the measured proton-proton reaction cross section 
($\sigma_{pp}^{\rm reaction}\equiv \sigma_{pp}^{\rm total} - 
\sigma_{pp}^{\rm elastic}$) is relatively small at $T_p$ less 
than $\sim1000$ MeV, where $\sigma_{pp}^{\rm total}$ 
($\sigma_{pp}^{\rm elastic}$) is the proton-proton total 
(elastic) cross section.  Once one accepts the 
scattering amplitude of the Fraunhofer scattering for 
describing the reactions~\cite{BS1}, one naturally obtains 
expression (\ref{rrf}) 
(see Sec.\ 2 of ref.~\cite{kurosupple}).
By substituting the values of $a$ determined by Eq.\ 
(\ref{form-a}) into Eq.\ (\ref{rrf}), we evaluate 
$\sigma_{\rm BS}$ for various nuclei at various proton energies.  

     The BS approximation is also applicable for analyzing 
nuclear matter radii.  For simplicity, we assume that 
the density distribution of the black sphere is uniform, 
i.e., a rectangular nucleon distribution.  Then 
we can naturally write the root-mean-square (rms) 
BS radius, $r_{\rm BS}$, as
\begin{equation}
   r_{\rm BS}\equiv \sqrt{3/5} a. 
   \label{rbb}
\end{equation}
The factor $\sqrt{3/5}$ comes from the second moment of 
the rectangular density distribution.  The values of 
$r_{\rm BS}$ are to be compared with the empirically deduced 
values of the rms matter radius, $r_m$, and in fact 
will turn out to be in good agreement with $r_m$ at 
$T_p\gtrsim 800$ MeV and $A\gtrsim50$.

\setcounter{equation}{0}
\section{How Do the Data Look Like?}
\label{how}

\begin{figure}[t]
\begin{center}
\includegraphics[width=7cm]{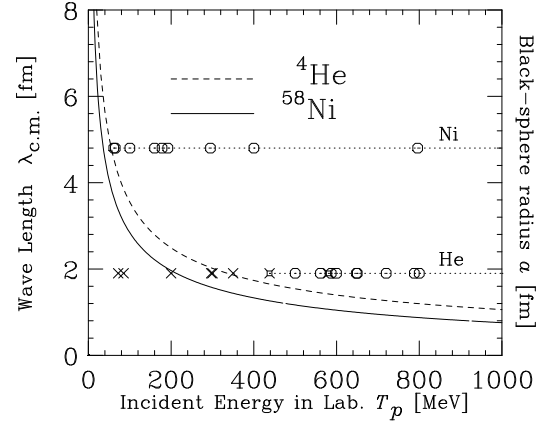} 
\end{center}
\vspace{-0.5cm}
\caption{Comparison of the BS radius, $a$, 
at $T_p \simeq$ 800 MeV with the c.m.\ de Broglie 
wave length of an incident proton of kinetic energy $T_p$
in the laboratory frame for the target of $^4$He ($^{58}$Ni).   
We draw the dashed (solid) curve for the de Broglie 
wave length for $^4$He ($^{58}$Ni). 
The dotted lines show the BS radius $a$ 
of $^4$He and $^{58}$Ni at about 800 MeV. 
On top of those lines, we plot the circles (crosses)
in the case in which the first peak of the diffraction 
appears (disappears).  The square with cross at 438 MeV 
for $^4$He implies that the first peak is not clear due 
to the quality of the data.} 
\label{peaks}
\end{figure}

     In this section, we overview proton scattering data 
and analyze them within the framework of
the BS approximation. 

\subsection{Resolution}
\label{reso}

     For validity of the BS approximation, the scattering 
should be close to the limit of the geometrical optics, 
as mentioned in the previous section.  This condition is 
fairly well satisfied at least for $T_p\gtrsim800$ MeV, since 
$a/\lambda_{\rm Lab}$ is well above unity even for $^4$He. 
The existence of the first diffraction peak is
a good measure to check the validity of the present
approximation as a function of energy. 
In fact, it is indispensable in our formulation 
to determine the value of $a$ from the empirical
diffraction peak angle of proton-nucleus elastic 
scattering.

     In order to examine the correspondence between the peak 
existence and the ratio of the BS radius $a$ at 
$T_p\gtrsim800$ MeV to the proton de Broglie wave length, 
$\lambda_{\rm c.m.}=2\pi/p$, in the c.m.\ frame,
we plot in Fig.~\ref{peaks} the ``won-and-lost records" 
which distinguish between the presence and absence of the
diffraction peak.  Note that the $T_p$ dependence of the BS 
radius is much weaker than that of $\lambda_{\rm c.m.}$ as we 
shall see.

     For $^4$He, we adopt the data of      
$T_p =72$ MeV~\cite{korshe:npa617}, 
$T_p =85$ MeV~\cite{votta}, 
$T_p =200, 350, 500$ MeV~\cite{moss},  
$T_p =297$ MeV~\cite{yoshimura}, 
$T_p =300$ MeV~\cite{yamagata}, 
$T_p =438, 648, 1036$ MeV~\cite{berger:prl37}, 
$T_p =350, 650, 1050, 1150$ MeV~\cite{aslanides}, 
$T_p =561, 800, 1029$ MeV~\cite{courant}, 
$T_p =580, 720$ MeV~\cite{verbeck}, 
$T_p =587$ MeV~\cite{boschitz}, 
$T_p =600$ MeV~\cite{fain}, 
and $T_p =788$ MeV~\cite{fong}. 
For $T_p \gtrsim800$ MeV, the references are listed 
in ref.~\cite{BS2}. 

     For $^{58}$Ni, we adopt the data of 
$T_p =61.4$ MeV~\cite{fulmer}, 
$T_p =65$ MeV~\cite{noro:npa,sakagu}, 
$T_p =100.4$ MeV~\cite{kwiatkowski}, 
$T_p =160$ MeV~\cite{roos}, 
$T_p =178$ MeV~\cite{ingemarsson:npa322},
and $T_p =192, 295, 400$ MeV~\cite{sakagu:prc57}.
For $T_p \gtrsim800$ MeV, the references are listed 
in ref.~\cite{BS1}. 

     As the incident energy decreases, the 
oscillation becomes broader and more blurred, and 
eventually the first peak disappears around 
$\lambda_{\rm c.m.}/a \simeq 1$, as shown in Fig.\ 
\ref{peaks}.  It is clearly shown in Fig.~2(a) of 
ref.~\cite{sakagu} how the measured diffraction peaks 
disappear at $T_p = 65$~MeV for $^{16}$O--$^{40}$Ca. 
At this energy, the de Broglie wave length is around 
4 fm, which is of order the values of $a$ at 800 MeV 
for such targets as $^{20}$Ne and $^{24}$Mg.  Thus, 
the presence of the diffraction peaks is closely 
related to the nuclear size.   As will be mentioned 
in the next subsection, the value of $a$ at $T_p \simeq 800$ 
MeV, multiplied by $\sqrt{3/5}$, is surprisingly 
close to $r_m$ for $A \gtrsim 50$. 

\subsection{$T_p\gtrsim800$ MeV}
\label{gt800}

     In refs.~\cite{BS1,BS2}, we clarified two salient
features from the data of incident energies higher than 
$\sim800$ MeV.  First, the absorption cross section
$\sigma_{\rm BS}$, Eq.~(\ref{rrf}), agrees with the 
empirical total reaction cross section within error bars,
i.e., 
\begin{equation}
   \sigma_R \simeq \sigma_{\rm BS}, 
\label{sigma-bs}
\end{equation}
although the comparison is possible only for stable nuclei
such as C, Sn, and Pb \cite{BS2}.  We will show later that this 
feature is persistent down to $T_p \sim100$ MeV.  We thus see 
the role played by $\sigma_{\rm BS}$ in predicting 
$\sigma_R$.  This is useful for nuclides for which elastic 
scattering data are available but no data for $\sigma_R$ 
are available.  Second, $r_{\rm BS}$, Eq.~(\ref{rbb}), 
almost completely agrees with the empirically deduced values 
of $r_m$ for $A\gtrsim50$, while it systematically 
deviates from the deduced values for $A\lesssim50$. 

     Let us examine the case of $A\lesssim50$ in more 
details. For light nuclei, the oscillation period 
in the elastic diffraction becomes broader as $A$ 
decreases~\cite{ADL}.  Equation~(\ref{form-a}) implies
that when $a$ becomes smaller, the value of $\theta_M$ 
becomes larger for fixed $T_p$.  Since the value of 
$\theta_M$ itself is relatively large,  
the values of $\sigma_{\rm BS}$ and $r_{\rm BS}$ can be well 
determined despite the uncertainty in $\theta_M$ as 
compared to heavy nuclei.  Therefore, better 
determination of $\sigma_{\rm BS}$ and $r_{\rm BS}$ would be 
possible for light nuclei than for heavy nuclei.

     In contrast to the good agreement between $r_{\rm BS}$ 
and $r_m$ for $A\gtrsim 50$, however, the values of 
$r_{\rm BS}$ are found to be systematically smaller than 
those of $r_m$ for $A\lesssim 50$.  A possible reason for
this discrepancy is the change in the ratio between the 
surface and the bulk portions toward lighter nuclei. 
This change may induce a difference in the expansion
series with respect to $A^{1/3}$ between $r_{\rm BS}$ and 
$r_m$, 
as discussed in Sec.~1.2
of Ref~~\cite{kurosupple}. 
We remark that the induced difference is appreciable
even for $A\gtrsim 50$, while the good agreement between
$r_{\rm BS}$ and $r_m$ suggests a counteracting effect 
due to developing neutron-skin thickness on the beta 
stability line \cite{kondo} 
(see also Sec.~4 of ref.~~\cite{kurosupple}).

\begin{figure}[t]
\begin{center}
\includegraphics[width=7cm]{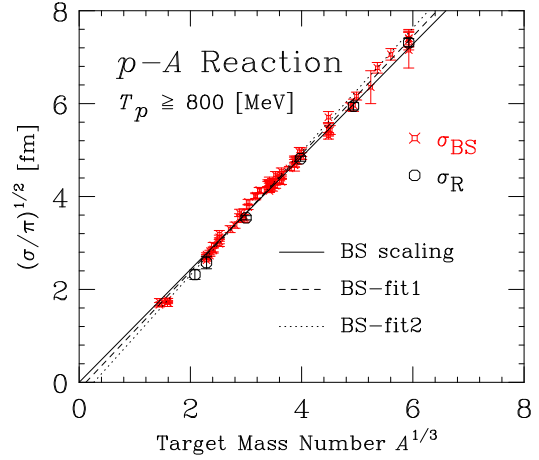} 
\end{center}
\vspace{-0.5cm}
\caption{(Color online)
Comparison of the three fitting lines with the
effective radius $\sqrt{\sigma_{\rm BS}/\pi}$ (crosses) for 
the absorption cross section of protons of 
$T_p\gtrsim800$ MeV by a target nucleus of mass number $A$.  
The solid line denotes the BS scaling, Eq.~(\ref{ascale}), 
the dashed line the BS-fit1, Eq.~(\ref{carlson1}), and the 
dotted line the BS-fit2, Eq.~(\ref{carlson2}).  
We also plot the effective radius 
$\sqrt{\sigma_R/\pi}$ (dots) for the empirical data 
for $^9$Be, $^{27}$Al, 
C, Cu, Sn, and Pb.  For the latter four elements the value of 
$\sigma_R$ is the average over the natural 
isotopic abundance in a target.  For these data, we set $A$ 
as the mass number of the most abundant isotope and assign 
the uncertainty in $A$ due to the natural abundance, 
as in Fig.~3 of ref.~\cite{BS2}. 
} 
\label{figscale}
\end{figure}

     From the observation of the global $A$ dependence
of the BS radius, $a$, we found that, for stable nuclei, 
the BS radius $a$ scales as~\cite{BS2}
\begin{equation}
  a \simeq 1.2135 A^{1/3} ~{\rm fm},
\label{ascale}
\end{equation}
which will be hereafter referred to as the black-sphere 
scaling (BS scaling).  Equivalently, from Eq.~(\ref{rbb}), 
we obtain~\cite{BS2}
\begin{equation}
  r_{\rm BS} \simeq 0.9400 A^{1/3} ~~{\rm fm},
\label{rbsscale}
\end{equation}
and, from Eq.~(\ref{rrf}), 
\begin{equation}
  \sigma_{\rm BS} \simeq 46.263 A^{2/3} ~~{\rm mb}. 
\label{sigbsscale}
\end{equation}
As one can see from Fig.~\ref{figscale}, the agreement of 
the BS scaling with both the empirical values of 
$\sqrt{\sigma_R/\pi}$ and $\sqrt{\sigma_{\rm BS}/\pi}$ 
is fairly good.

     For systematic evaluations of $\sigma_R$ that
can be applied for practical use, we have to aim at a better 
agreement with the empirical values beyond expression 
(\ref{sigbsscale}).  For this purpose, we propose a couple of 
other parametrizations \`{a} la Carlson, which fit 
$(\sigma_R/\pi)^{1/2}$ in terms of a linear function in 
$A^{1/3}$~\cite{carlson}:  The first one, denoted by 
BS-fit1, is given by 
\begin{equation}
   a = 1.2671A^{1/3} - 0.152~{\rm fm}.  
\label{carlson1}
\end{equation}
We obtain this by $\chi^2$-fitting of the linear function in 
$A^{1/3}$ to the values of $(\sigma_{\rm BS}/\pi)^{1/2}$. 
The standard deviation is around 0.096 fm.   Note that    
this expression inevitably puts its emphasis on the region of 
$A<50$, because most of the data points distribute in 
this region.  By putting slightly more emphasis on the 
data points in the region of $A>50$, we obtain the 
second one, BS-fit2, which is given by 
\begin{equation}
   a = 1.33A^{1/3} - 0.35~{\rm fm}. 
\label{carlson2}
\end{equation}
Note that in Fig.\ \ref{figscale} 
the BS-fit1 intervenes between the BS scaling
and the BS-fit2, 
which differ only by of order 0.1 fm.

\subsection{Down to $T_p\simeq50$ MeV}

\begin{figure}[t]
\begin{center}
\includegraphics[width=7cm]{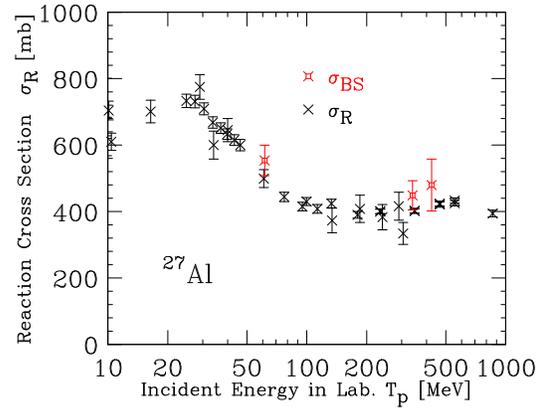} 
\end{center}
\vspace{-0.5cm}
\caption{(Color online) 
Energy dependence of $\sigma_{\rm BS}$ (squares with cross) 
for the reaction of protons on $^{27}$Al. 
The values of $\sigma_{\rm BS}$ 
are obtained from the measured peak angle of the first 
diffraction maximum of the proton elastic scattering. 
The empirical data for $\sigma_R$ ($\times$) 
are taken from in ref.~\cite{carlson}. 
} 
\label{edepbs}
\end{figure}

     Let us consider how the data look like when we 
decrease $T_p$ from 800 MeV.  As we have mentioned 
in Sec.~\ref{reso} for a target of fixed $A$, the diffraction 
patterns become blurred as $T_p$ decreases. 
At a certain value of $T_p$, the first diffraction peak 
tends to disappear.  We find, however, that, 
as long as the peak exists, the relation~(\ref{sigma-bs}) 
holds within the empirical uncertainties.

\begin{figure}[t]
\begin{center}
\includegraphics[width=7cm]{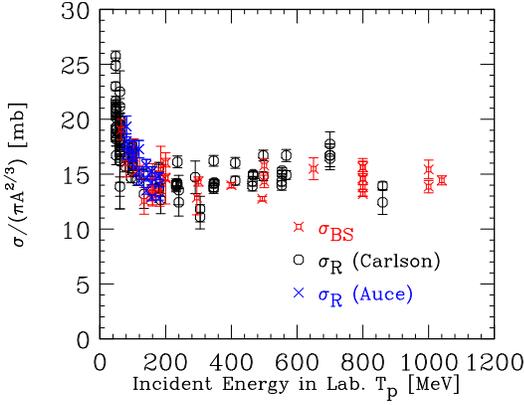} 
\end{center}
\vspace{-0.5cm}
\caption{(Color online) Comparison of the effective cross section
$\sigma_{\rm BS}/(\pi A^{2/3})$ with $\sigma_R/(\pi A^{2/3})$ 
as a function of the kinetic energy of an incident proton. 
The empirical data for $\sigma_R$ are taken 
from the compilation by Carlson ($\circ$)~\cite{carlson} 
and the measurements by Auce {\it et al}.\ ($\times$)~\cite{auce}. 
The values of $\sigma_{\rm BS}$ ($\equiv \pi a^2$), 
which are represented by squares with crosses,  
are obtained from the measured peak angle of the first diffraction 
maximum of the proton elastic scattering. They are consistent 
with the measured $\sigma_R$. 
} 
\label{edep-effxec}
\end{figure}

     As an example, in Fig.~\ref{edepbs}, we plot 
the values of $\sigma_{\rm BS}$ for proton-$^{27}$Al
reactions as a function of $T_p$.  To obtain the 
values of $a$, we adopt the first peak angle of the 
empirical data for the differential cross sections 
of proton-$^{27}$Al elastic scattering at $T_p =61.4$ 
MeV~\cite{fulmer}, 340 MeV~\cite{richardson}, and 
424 MeV~\cite{heibeg}.  Other examples will be 
shown later.  From Fig.~\ref{edepbs}, we find that
$\sigma_R$ agrees with $\sigma_{\rm BS}$ within error bars,
which ensures Eq.~(\ref{sigma-bs}).  This relation suggests 
that the radius $a$ can be regarded as a ``reaction radius,'' 
inside which the reaction with incident protons occurs.
The above tendency holds for other stable nuclei,
as shown in Fig.~\ref{edep-effxec}.

     Note that the condition for Eq.~(\ref{sigma-bs})
to hold is slightly different for the case of nucleus-nucleus 
reactions.  As we have already shown in refs.~\cite{BS3,kio08}, 
the BS approximation can be extended to nucleus-nucleus reactions 
by using $\pi(a_P + a_T)^2$, where $a_P$ ($a_T$) is the BS 
radius of a projectile (target).  Interestingly, the 
empirical values of the total reaction cross section $\sigma_R(A+A)$ 
agree well with $\pi(a_P+a_T)^2$ for incident energies per
nucleon down to 100 MeV, not only in the presence of the 
first diffraction peaks of proton elastic scattering that
lead to $a_P$ and $a_T$, but also in their absence in which case
$a_P$ and $a_T$ are evaluated from Eq.\ (\ref{sigma-bs}) as
$(\sigma_R/\pi)^{1/2}$.

\setcounter{equation}{0}
\section{Black-sphere (BS) Cross-Section Formula} 
\label{bsxsec}

\begin{figure}[t]
\begin{center}
\hspace*{-0.65cm}
\includegraphics[width=9.5cm]{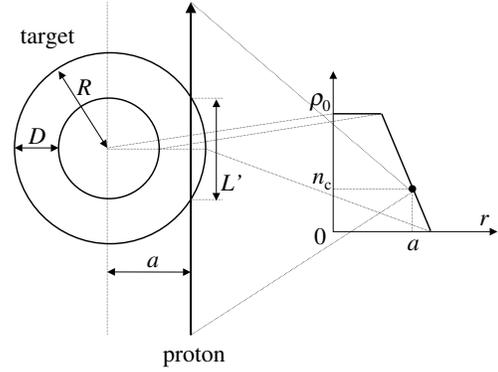} 
\end{center}
\vspace*{-0.5cm}
\caption{Model for the density distribution of a target nucleus and the 
critical proton trajectory inside which the reaction with the target
nucleus occurs. This is the same as Fig.~1 in ref.~\cite{BS3}. 
} 
\label{trapezoidal}
\end{figure}

     In this section, we briefly review the BS cross-section 
formula, which was originally developed in ref.~\cite{BS3}
for describing the proton-nucleus total reaction cross 
section $\sigma_R$, and analytically analyze the $A$ 
dependence of the formula.  The detailed derivation of 
the formula 
can be found in Sec.\ 3 of ref.~\cite{kurosupple}.

     The BS cross-section formula is constructed as a 
function of the mass and neutron excess of the target nucleus 
and $T_p$ in a way free from any adjustable $T_p$-dependent 
parameter.  
The geometry of the reaction is assumed 
as can be seen in Fig.~\ref{trapezoidal}.
We deduce the dependence of $\sigma_R$ 
on $T_p$ from a simple argument involving the nuclear 
``optical" depth for absorption of incident protons within 
the framework of the BS approximation of nuclei.  This 
formula can be easily extended to nucleus-nucleus 
reactions~\cite{BS3,kio08}.

\subsection{Energy Dependence}
\label{sec-edep}

\begin{figure}[t]
\begin{center}
\includegraphics[width=7cm]{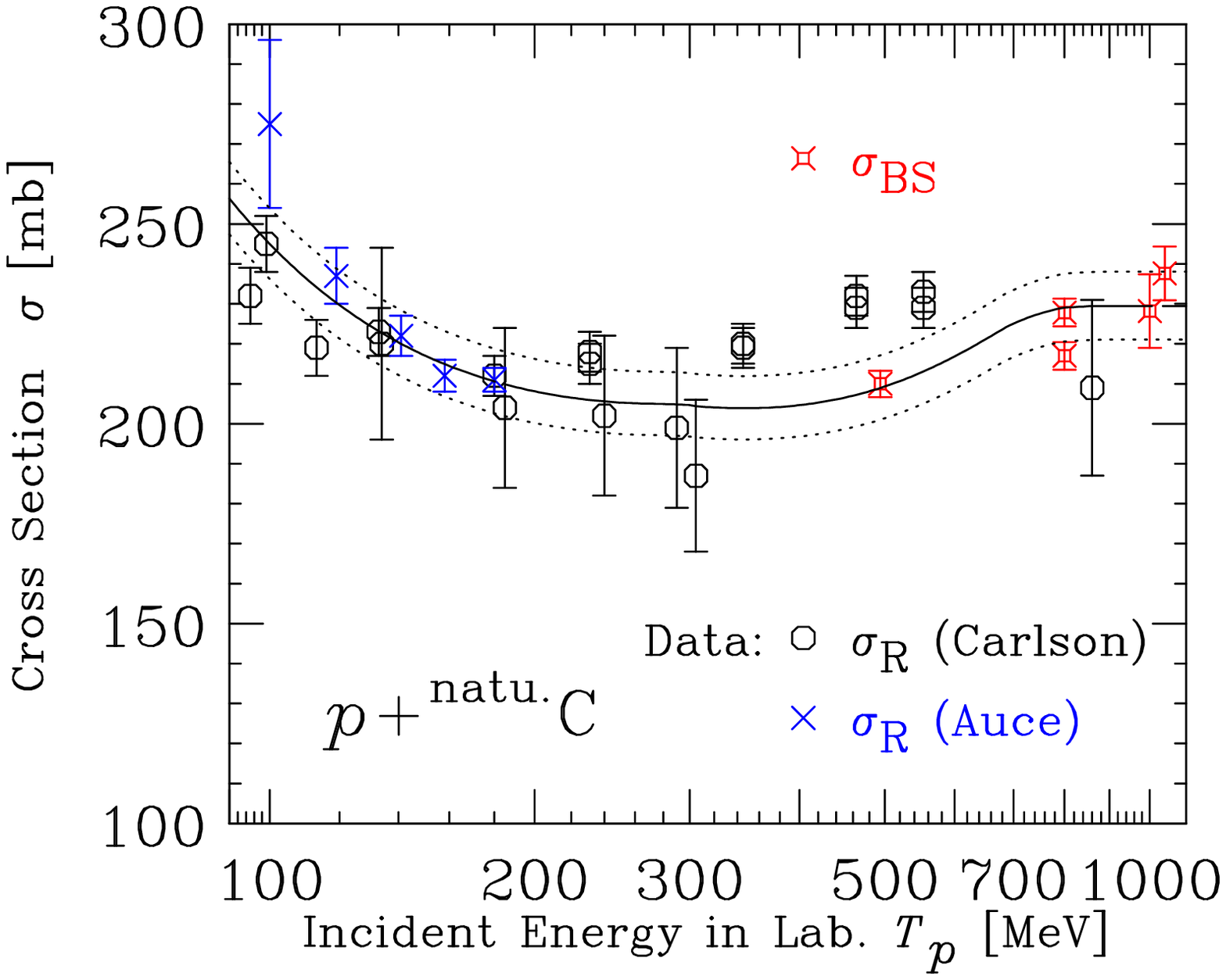} 
\includegraphics[width=7cm]{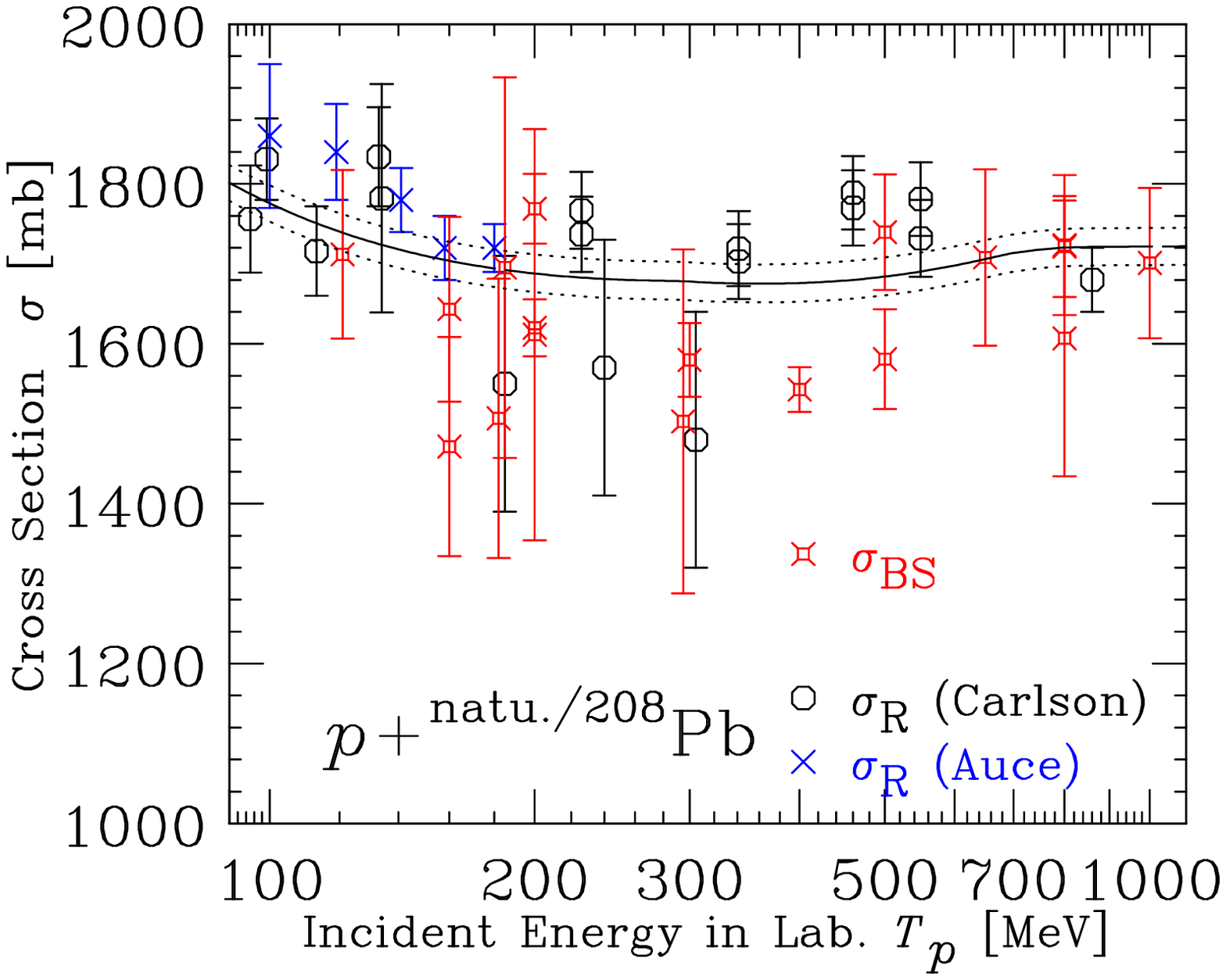} 
\end{center}
\vspace{-0.5cm}
\caption{(Color online) Comparison of the BS cross-section 
formula (solid curve) with the empirical values for 
$\sigma_R(p+{^{\rm natu.}{\rm C}})$ (upper)
and $\sigma_R(p+{^{\rm natu./208}{\rm Pb}})$ (lower)
as a function of the kinetic energy of an incident proton. 
We adopt the BS radius, $a_0$, at 800 MeV 
as $2.70\pm0.05$ fm for carbon and $7.40\pm0.05$ fm for lead. 
The uncertainties, whose assignment is the same 
as in ref.~\cite{BS1}, are shown by the dotted curves. 
We also plot the empirical data for $\sigma_R$ 
from the compilation by Carlson ($\circ$)~\cite{carlson} 
and the measurements by Auce {\it et al}.\ ($\times$)~\cite{auce}. 
Note that the data of 220 MeV $\le T_p \le 570$ 
MeV~\cite{ren:npa} turn out to be systematically large. 
The values of $\sigma_{\rm BS}$ ($\equiv \pi a^2$), 
which are represented by squares with crosses, 
are obtained from the measured peak angle of the first diffraction 
maximum of the proton elastic scattering.  They are consistent 
with the measured $\sigma_R$. 
} 
\label{edep-cpb}
\end{figure}

     In setting the $T_p$ dependence of the formula, we retain
the expression for $\sigma_{\rm BS}$ given by Eq.~(\ref{rrf}), 
leading to
\begin{eqnarray}
  \tilde{\sigma}_{\rm BS}(T_p) 
  &=& \pi a(T_p)^2
\nonumber\\
  &=& \pi a_0^2 \left(1 + {\Delta a \over a_0} \right)^2, 
\label{form1a}
\end{eqnarray}
where $\Delta a \equiv a(T_p) - a_0$, and $a_0$ denotes the 
value of $a$ determined at 800 MeV for each nucleus.  While
Eqs.~(\ref{ascale}), (\ref{carlson1}), and (\ref{carlson2}) 
introduced in Sec.\ \ref{gt800} are helpful for estimating $a_0$,
$\Delta a$ is responsible for the energy dependence of $\sigma_{\rm BS}$. 
We introduce the {\it effective} nuclear optical depth 
$\tau$ defined by 
\begin{equation}
  \tau = \bar{\sigma}_{pN}^{\rm total} n_c L', 
\label{efftau}
\end{equation}
with
\begin{equation}
  \bar{\sigma}_{pN}^{\rm total} 
  = (Z/A) \sigma_{pp}^{\rm total} + (1 - Z/A) \sigma_{pn}^{\rm total}, 
\label{sigtot-ave}
\end{equation}
where $\sigma_{pp(pn)}^{\rm total}$ is the proton-proton (neutron) 
total cross section, 
$n_c$ is the critical nucleon density
at the distance of $r = a$ from the nuclear center,
and $L'$ is the length of the part of the critical 
trajectory in which the total nucleon density is lower than $n_c$ 
while being above zero 
(see Fig.~\ref{trapezoidal}).

     By assuming that $\tau= 0.9$ independently of $T_p$, 
we express the $T_p$ dependence of $\Delta a/a_0$ as that driven 
solely by $\bar{\sigma}_{pN}^{\rm total}$.  The 
$T_p$-$in$dependent part of $\Delta a/a_0$ is described 
by several parameters that characterize the density distribution 
of the target nucleus assumed to be trapezoidal, which makes the 
expression for $\Delta a$ analytically tractable.  The choice of the 
value of 0.9 for $\tau$ is reasonable since this is consistent with the 
values of $a_0$ and $n_c$ for $^{12}$C, $^{58}$Ni, $^{124}$Sn, and 
$^{208}$Pb listed 
in Table~1 of ref.~\cite{kurosupple}.  
The detailed description of the formula is given 
in the original paper~\cite{BS3} and 
in 
Sec.\ 3 of ref.~\cite{kurosupple} 
in which minor corrections
to the $T_p$-independent part of 
$\Delta a/a_0$ described in ref.~\cite{BS3} will be added.

     The comparison with the empirical data is shown in 
Fig.~\ref{edep-cpb}.  We find that, for stable nuclei, this formula 
remarkably well reproduces the empirical $T_p$ dependence of 
$\sigma_R$ at $T_p=100$--1000 MeV, where the deviation of 
$\bar{\sigma}_{pN}^{\rm total}$ from its empirical value at 
$T_p =800$ MeV is small enough to validate the present formulation. 
We remark in passing that the contribution from the Coulomb 
interaction, which is not included in this framework, 
can be safely neglected in this energy region~\cite{BS3}.

\begin{figure}[t]
\begin{center}
\includegraphics[width=7cm]{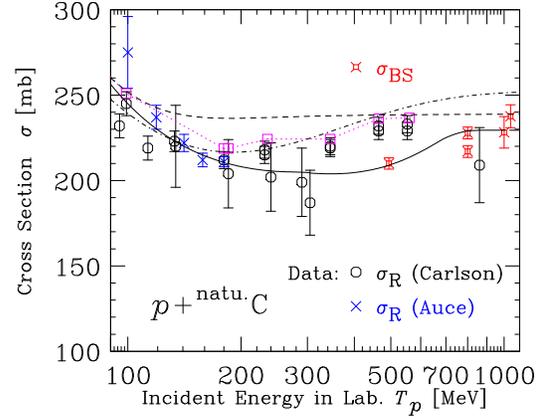} 
\end{center}
\vspace{-0.5cm}
\caption{(Color online) Comparison of various empirical 
formulas for $\sigma_R(p+{\rm C})$
with the empirical values as a function of $T_p$.
The values of Carlson's formula obtained by using the 
parameters listed in Table~A of ref.\ \cite{carlson} 
are plotted by 
the squares with dotted line.  The values 
of the same formula but using the parametrization of Machner 
{\it et al}.\ \cite{machner} are drawn by the dashed curve.  
The values of Letaw's formula \cite{letaw}
are drawn by the dot-dashed curve.  The 
values of the BS cross-section formula with $a_0 = 2.70$ fm 
are drawn by the solid curve.  The empirical values and the 
values of $\sigma_{\rm BS}$ are the same as in 
Fig.~\ref{edep-cpb}. 
} 
\label{empfig}
\end{figure}

     We also plot the values of $\sigma_{\rm BS}$ in the figure. 
For obtaining these values, we adopt the empirical data
for the differential cross sections of proton-C elastic scattering 
at $T_p = 494.0$ MeV in ref.~\cite{hoffmann:prc41}.  For 
$T_p \gtrsim800$ MeV, the references of the data are listed in 
ref.~\cite{BS2}.  For elastic scattering data of $^{208}$Pb,   
we adopt the empirical values at $T_p = 65$ MeV~\cite{sakagu}, 
$T_p =80, 121, 160, 182$ MeV~\cite{nadasen}, 
$T_p =160$ MeV~\cite{roos},  
$T_p =185$ MeV~\cite{oers}, 
$T_p =200$ MeV~\cite{mcdaniels,lee:plb205}, 
$T_p =200, 300, 400, 500$ MeV~\cite{hutcheon:npa483}, 
$T_p =295$ MeV~\cite{zeni}, 
$T_p =500$ MeV~\cite{hoff:prl47}, and 
$T_p =650$ MeV~\cite{mack:prc52}.
We do not adopt the data at $T_p = 100.4$ MeV~\cite{kwiatkowski}, 
because the measured diffraction pattern does not 
include the first peak.  For $T_p \gtrsim 800$ MeV, the references 
are listed for $p+^{208}$Pb in ref.\ \cite{BS1}. 

     For comparison, we plot the present BS cross-section
formula together with other empirical formulas in 
Fig.~\ref{empfig}.  We choose those constructed only for 
nucleon-nucleus reactions, which are summarized in 
Sec.~6 of ref.~\cite{kurosupple}.
The present formula alone 
reproduces the $T_p$ dependence in a manner that is 
consistent with the latest empirical data~\cite{auce}, which are 
systematically more reliable.

\subsection{Mass-Number ($A$) Dependence}
\label{bs-adep}

This section provides a main part of the present paper. 
     Here, we examine the $A$ dependence of the BS cross-section 
formula, 
\begin{eqnarray}
  \tilde{\sigma}_{\rm BS}(T_p) 
  = \pi a_0^2 
    \left[1 + \left(\frac{\rho_0 a_0}{D n_{c0}}
                  - {a_0 \over L'_0} 
                    \left. {dL' \over da}\right|_0 
              \right)^{-1}
              \frac{\Delta\bar{\sigma}_{pN}^{\rm total}}
            {\bar{\sigma}_{pN0}^{\rm total}} 
    \right]^2.   
\nonumber\\
\label{form0}
\end{eqnarray}
The definition of each term in this formula can be found in
Sec.\ 3 of ref.~\cite{kurosupple},
together with the derivation of Eq.~(\ref{form0})
(or, equivalently, Eq. (3.12) of ref.~\cite{kurosupple}).

     First, we examine the $A$ dependence of the key terms 
     in the above expression. 
Hereafter, just like $a_0$, we will affix ``0'' to the $T_p$ 
dependent quantities whenever we mean the values at $T_p=800$ MeV. 
Following the definition of $L'$, 
Eq.~(3.4) of ref.~\cite{kurosupple},
the $A$ dependence of the path length, $L'$, can be expressed 
as 
\begin{eqnarray}
   L' &=& 2\sqrt{R^2 - a^2}
\nonumber\\
      &=& 2\sqrt{(R + a)(R - a)}
\nonumber\\
      &\propto& A^{1/6} D^{1/2}, 
\label{optdep}
\end{eqnarray}
where $D(= 2.2 {\rm fm})$ is a constant as given 
in Eq.~(3.10) of ref.~\cite{kurosupple}.

Another key term is $\rho_0 a_0/(D n_{c0})$ in the coefficient 
of $\bar{\sigma}_{pN}$ found in expression~(\ref{form0}).  Since 
$a_0 \propto A^{1/3}$ and 
$n_{c0} \propto L'^{-1}\propto A^{-1/6} D^{-1/2}$, we may write 
\begin{equation}
   \frac{\rho_0 a_0}{D n_{c0}} \propto D^{-1/2} A^{1/2}, 
\label{optdeplead}
\end{equation}
where $\rho_0= 0.16$ fm$^{-3}$. 

     We next examine the coefficient of 
$\Delta\bar{\sigma}_{pN}^{\rm total}/\bar{\sigma}_{pN0}^{\rm total}$,
i.e., 
\begin{equation}
\left(\frac{\rho_0 a_0}{D n_{c0}}- {a_0 \over L'_0} 
                    \left. {dL' \over da}\right|_0 
              \right)^{-1},    
\label{coeff1}
\end{equation}
in Eq.~(\ref{form0}).  
To examine the $A$ dependence of the second term 
in the parenthesis of Eq.~(\ref{coeff1}), 
we look into the term 
$dL'/da|_0$ 
(see Eq.~(3.19) of ref.~\cite{kurosupple}).
We obtain
\begin{eqnarray}
   &{}&  \frac{\rho_0 a_0}{D n_{c0}}
      - \frac{a_0}{L'_0} \left. \frac{dL'}{da}\right|_0 
    \nonumber\\
   &=& \frac{\rho_0 a_0}{D n_{c0}}
       \left\{1 + 
             4 \frac{D n_{c0}}{\rho_0 a_0}
             \left(\frac{a_0}{L_0'}\right)^2 
        \right\}.                   
\end{eqnarray}
Since the first term in the left side is much larger 
than the second term, 
we expand it as follows: 
\begin{eqnarray}
   &{}& \left(\frac{\rho_0 a_0}{D n_{c0}}
      - \frac{a_0}{L'_0} \left. \frac{dL'}{da}\right|_0 \right)^{-1}
\nonumber\\
   &=& \frac{D n_{c0}}{\rho_0 a_0}
     - 4 \left(\frac{D n_{c0}}{\rho_0 a_0}\right)^2
         \left(\frac{a_0}{L_0'}\right)^2 
         + \cdots.
\end{eqnarray}
This expansion becomes better as $A$ increases.  Thus, 
the first term gives the leading correction to 
$\pi a_0^2$ in the BS cross-section formula.

    Finally, we obtain the following expression: 
\begin{eqnarray}
  &{}& \tilde{\sigma}_{\rm BS}(T_p)
\nonumber\\
  &\simeq& \pi a_0^2 
  +  2\pi \left[\frac{D n_{c0}}{\rho_0 a_0} a_0^2 
+ O\left(\left(\frac{D n_{c0} a_0}{\rho_0 L_0'}\right)^2
              \right) 
              \right]
        \frac{\Delta\bar{\sigma}_{pN}^{\rm total}}
            {\bar{\sigma}_{pN0}^{\rm total}} 
\nonumber\\
  & & + O\left(
          \left(\frac{Dn_{c0}}{\rho_0}\right)^2 
        \left(\frac{\Delta\bar{\sigma}_{pN}^{\rm total}}
             {\bar{\sigma}_{pN0}^{\rm total}}\right)^2
        \right).
\label{form-adep}
\end{eqnarray}
Since, from Eq.~(\ref{optdeplead}), 
$D n_{c0}/(\rho_0 a_0)$ $\propto D^{1/2} A^{-1/2}$ and 
$a_0^2$ $\propto A^{2/3}$, we find that 
in the subleading term, 
$D n_{c0}/(\rho_0 a_0) a_0^2$ is proportional to $D^{1/2}A^{1/6}$.  
In this way, we analytically find that, in contrast to other 
formulas, our formula includes the $O(A^{1/6})$ term in addition 
to the leading $O(A^{2/3})$ 
term in $\tilde{\sigma}_{\rm BS}(T_p)$. 
The 
presence of the $O(A^{1/6})$ term, which comes from the nuclear 
optical depth, is 
{\em one of the salient features of the present 
formula}.

\begin{figure}[t]
\begin{center}
\includegraphics[width=7cm]{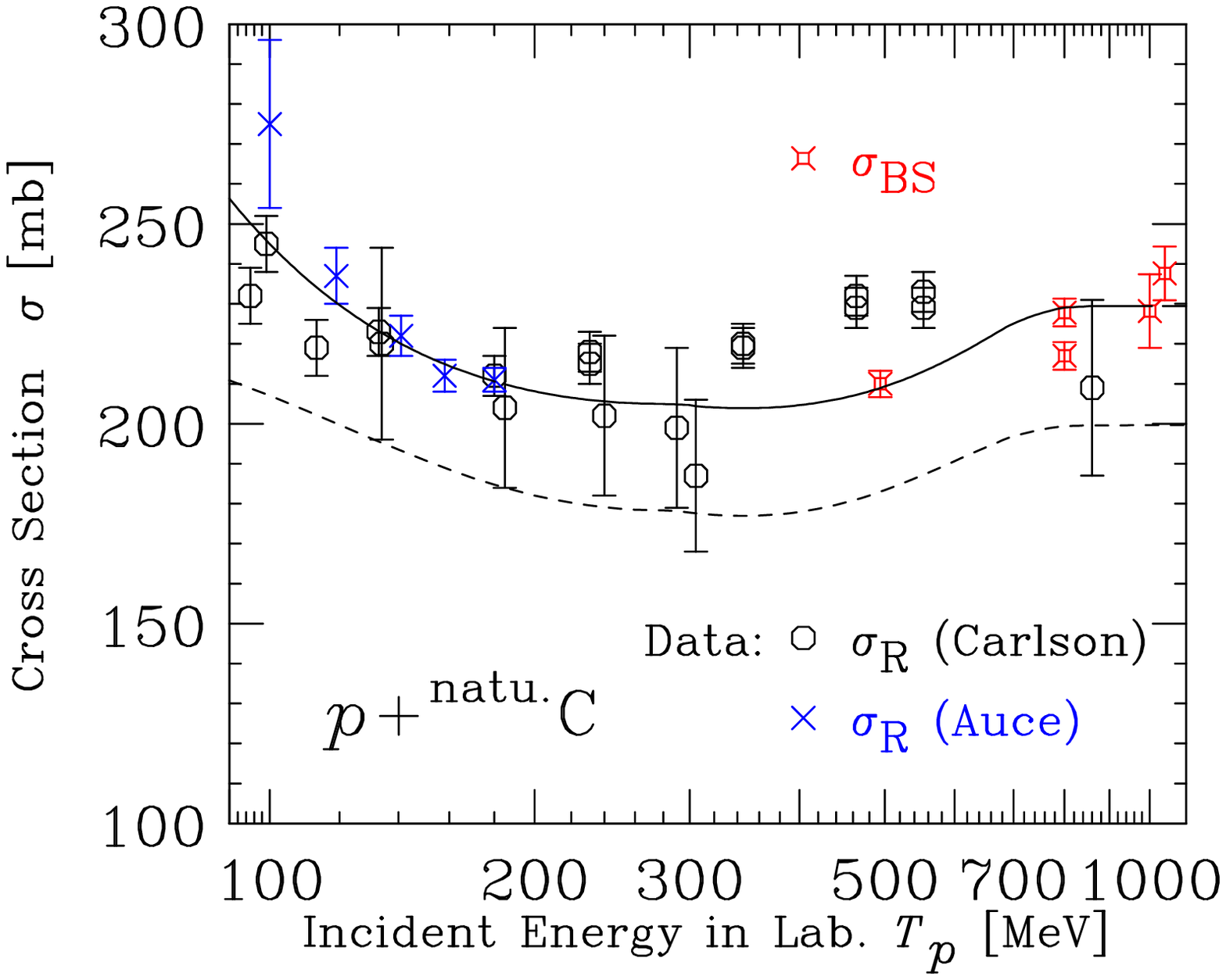}  
\includegraphics[width=7cm]{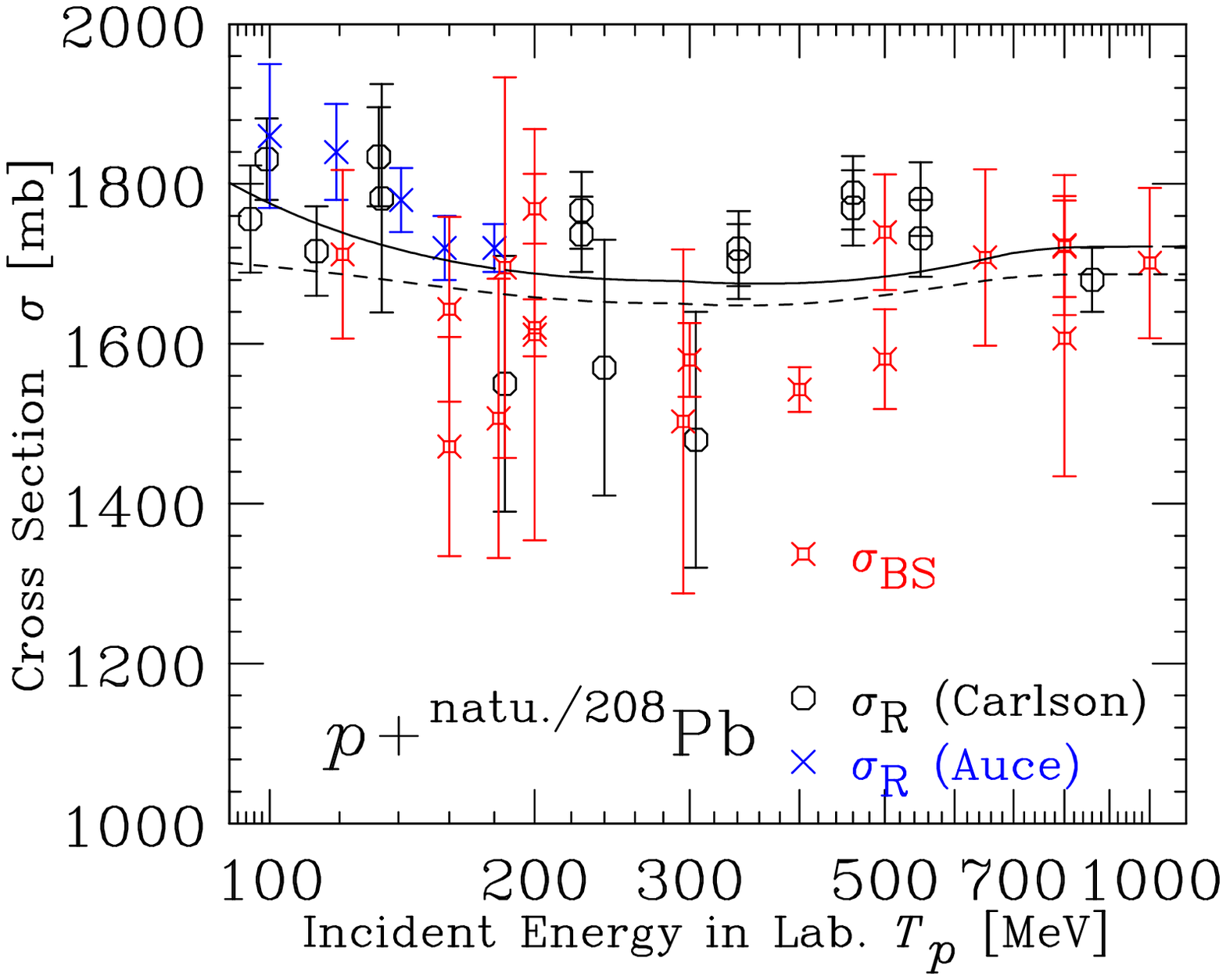}  
\end{center}
\vspace{-0.5cm}
\caption{(Color online) Comparison of the BS cross-section 
formula (solid curve) with the eikonal approximation based on 
the square-well potential (dashed curve) for 
$\sigma_R(p+{^{\rm natu.}{\rm C}})$ (upper) and 
$\sigma_R(p+^{\rm natu./208}{\rm Pb})$ (lower) 
as a function of the 
kinetic energy of an incident proton.  We adopt both the 
BS radius at 800 MeV and the square-well radius as 2.70 fm 
for carbon and 7.40 fm for lead.  
The empirical data for $\sigma_R$ and the values of 
$\sigma_{\rm BS}$ are the same as in Fig.~\ref{edep-cpb}.
} 
\label{figsqwell}
\end{figure}

     In order to illustrate the contribution from the $O(A^{1/6})$ term, 
in Fig.~\ref{figsqwell}, we compare the values of the BS 
cross-section formula (solid curve) with the values obtained by using 
the square-well potential 
within the eikonal approximation 
(dashed curve) for the cases of $^{\rm natu.}$C and Pb. 
The expression for $\sigma_R$ 
in the eikonal approximation
can be obtained from 
the square-well potential as Eq.~(\ref{g-sigabs4}) in  
Appendix~\ref{sec-sqwell}.  In this expression, for simplicity, 
we do not distinguish between protons and neutrons in the 
target.  As a result of expansion, the leading term is 
proportional to $A^{2/3}$, while the subleading term is 
proportional to $A^{1/3}$ multiplied by an $A$ dependent exponential 
suppression factor 
as can be found in 
Eq.~(\ref{g-sigabsadep}), 
which causes a different $T_p$ dependence from
the solid curve in each panel of Fig.~\ref{figsqwell}. 

     By comparing the solid curves in the upper and lower panels 
of Fig.~\ref{figsqwell}, one can see the relatively weaker 
$T_p$ dependence for the case of Pb.  The cross section itself 
grows proportional to $\sim A^{2/3}$, while the 
$T_p$-dependent term is proportional to $\sim A^{1/6}$, 
leading to $O(A^{-1/2})$ corrections to the $O(A^{2/3})$ term.
Thus, the relative change in the cross section by $T_p$ 
is suppressed.  This is the reason why the slope toward a
lower $T_p$ becomes steeper for the case of C than that of Pb. 
The latest empirical values of $\sigma_R$ 
~\cite{auce} apparently support the presence of 
the $T_p$-dependent $O(A^{-1/2})$ corrections.

     Furthermore, both for C and Pb, the BS cross-section 
formula shows a stronger $T_p$ dependence at 
$T_p \lesssim 200$ MeV than the case of the square-well 
potential.  This is because of the different dependence 
of the subleading term on $\sigma^{\rm total}_{pN}$.  
The former is 
the positive power law, while the latter is the negative 
power law multiplied by an exponential suppression factor
as in Eq.\ (\ref{g-sigabs4}).  This exponential factor 
drastically reduces the $T_p$ dependence of expression
(\ref{g-sigabs4}), which 
is at odds with a power-law $\sigma^{\rm total}_{pN}$ 
dependence of the reaction cross section that is empirically 
suggested \cite{IL}.

     In such conventional multiple-scattering theory as 
the Glauber approximation, even if one adopts a realistic 
density distribution as an input, a similarly weak $T_p$ 
dependence of the calculated $\sigma_R$ is 
suggested as shown in 
Fig.~2 
of ref.~\cite{bhks}.  The 
influence of the surface diffuseness is secondary in this 
context.  Therefore, a simplified comparison with our formula 
using the rectangular density distribution 
in the eikonal approximation makes sense for the purpose of
clarifying the essential difference between the two 
approaches.

\subsection{Overestimation by the BS Cross-Section Formula 
at Low Energy}
\label{overestimate}

\begin{figure}[t]
\begin{center}
\includegraphics[width=7cm]{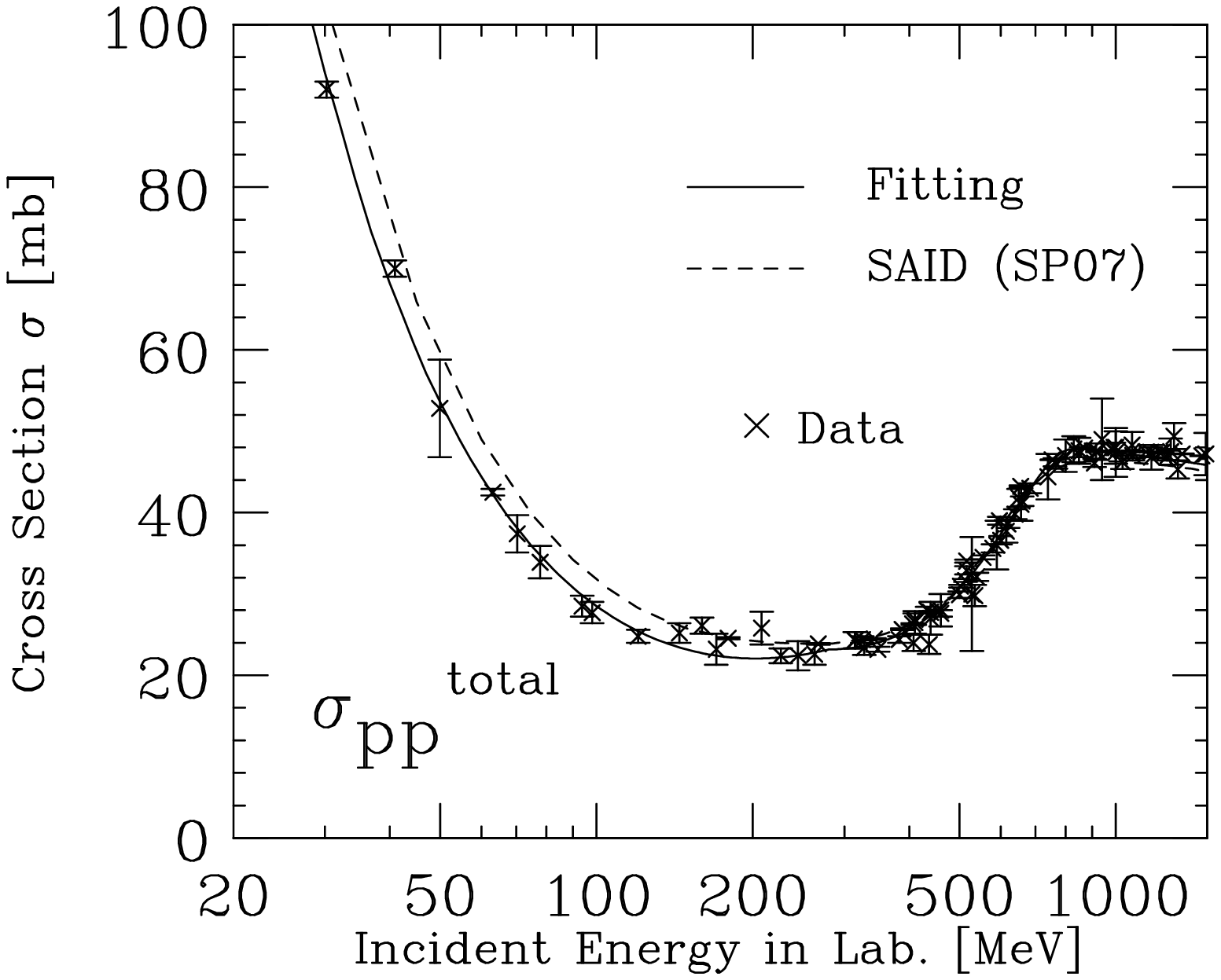} 
\includegraphics[width=7cm]{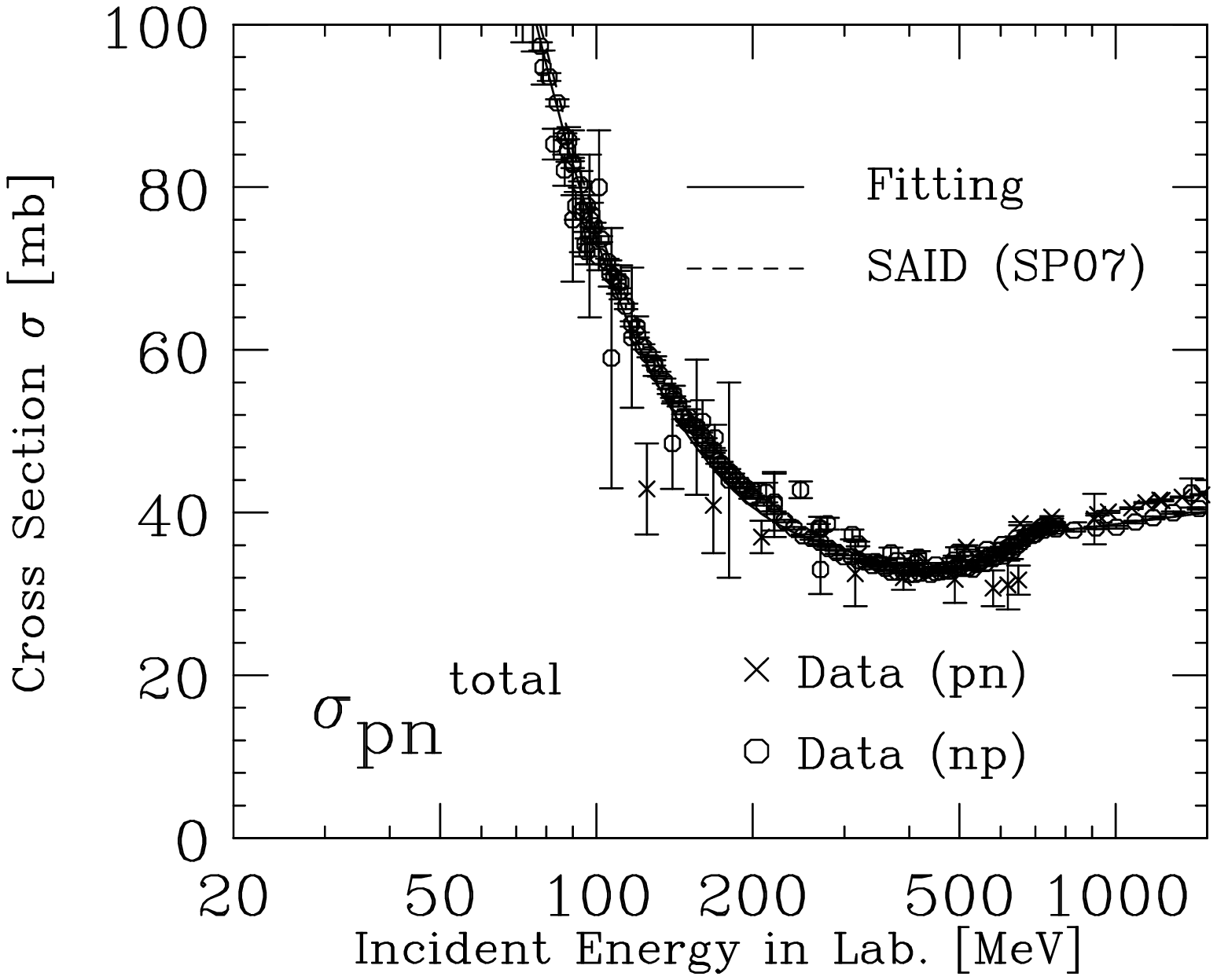} 
\end{center}
\vspace{-0.5cm}
\caption{ The empirical values of $\sigma_{pN}^{\rm total}$ 
as a function of nucleon incident energies lower
than 1.5 GeV.  The upper panel shows $\sigma_{pp}^{\rm total}$
and the lower panel $\sigma_{pn}^{\rm total}$. 
These values are obtained from the compilation 
by the Particle Data Group~\cite{pdg}. 
What the solid and dashed curves stand for 
can be found in Appendix~\ref{sigmatotal}. 
} 
\label{figsigpn}
\end{figure}

     At $T_p \lesssim 100$ MeV, the values of the BS 
cross-section formula overestimate the measured values of
$\sigma_R$.  We discuss this fact briefly in this subsection. 

     In constructing the formula, we assume that 
$|\sigma_{pN}^{\rm total}-\sigma_{pN0}^{\rm total}|$ is 
sufficiently small to validate $|\Delta a| \ll a_0$, which is 
well satisfied for $T_p\gtrsim 100$ MeV.  In this sense, the 
overestimation by the formula simply suggests that 
the approximation that we adopted becomes invalid.

     Let us look at which energy the value of 
$\sigma_{pN}^{\rm total}$ is equal to that of 
$T_p\simeq 800$ MeV, which we denote by 
$\sigma_{pN0}^{\rm total}$ 
in Sec.\ 3 of ref.~\cite{kurosupple}.
From Fig.~\ref{figsigpn}, 
which plots $\sigma_{pN}^{\rm total}$ as a function of 
the kinetic energy of an incident nucleon, one can observe 
that $\sigma_{pp}^{\rm total} \simeq \sigma_{pp0}^{\rm total}$ 
at $T_p\simeq 50$ MeV and $\sigma_{pn}^{\rm total} \simeq 
\sigma_{pn0}^{\rm total}$ at $T_p\simeq 200$ MeV. 
Since we adopt the averaged value in the formula 
as shown in Eq.~(\ref{sigtot-ave}),  
the relevant quantity is $\sigma_{pN}^{\rm total}$, and 
$\sigma_{pN}^{\rm total} \simeq \sigma_{pN0}^{\rm total}$ 
at $T_p\simeq 120$ MeV for $N = Z$.  Consequently,
the value of the BS cross-section formula for $p+A$ 
around $T_p=110$ MeV becomes the same as that of 
$T_p= 800$ MeV.  At $T_p\sim 110$ MeV, the formula starts
to deviate from the empirical values, and the deviation 
increases drastically as $T_p$ decreases, because 
$|\sigma_{pN}^{\rm total}-\sigma_{pN0}^{\rm total}|$ 
becomes too large for $|\Delta a| \ll a_0$ to be satisfied.

\setcounter{equation}{0}
\section{Other Probes}
\label{other-probes}

\begin{figure}[t]
\begin{center}
\includegraphics[width=7cm]{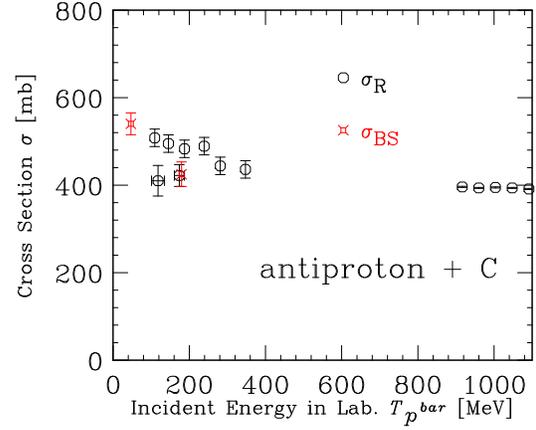} 
\end{center}
\vspace{-0.5cm}
\caption{(Color online) Comparison of $\sigma_{\rm BS}$ 
(squares with cross) with the empirical
$\sigma_R$ (circles) for antiproton-nucleus 
reactions as a function of incident kinetic energy 
$T_{\bar p}$ of antiprotons.  For calculating the values of 
$\sigma_{\rm BS}$, we adopt the empirical values of the angular 
distributions for elastic scattering from $^{12}$C at 
46.8 MeV~\cite{garreta1} and 179.7 MeV~\cite{garreta2}. 
For comparison, we plot the empirical values of $\sigma_R$ 
for a C target at the incident momenta of 466-879 MeV/$c$ 
($T_{\bar p} =$ 109.3, 145.3, 187.4, 239.0, 281.2, and 347.4 MeV)
~\cite{nakamurak}, 485 and 597 MeV/$c$ ($T_{\bar p} =$ 117.9 and 
173.8 MeV)~\cite{aihara}, and 1.6-1.8 GeV/$c$ ($T_{\bar p} =$ 
916.6, 959.9, 1003.5, 1047.4, and 1091.6 MeV)~\cite{abrams}. 
The above empirical values at $T_{\bar p} =$ 117.9 and 173.8 MeV 
are plotted with the uncertainties of around several percent in 
$T_{\bar p}$.  
} 
\label{fig-pbar}
\end{figure}

     It is natural to attempt to extend the BS 
approximation of nuclei to the processes of other hadronic probes 
such as antiprotons, pions, and kaons. 
Although the BS approximation is originally expected 
to provide a decent description of the reaction cross sections for 
any kind of incident particle that tends to be attenuated in nuclear 
interiors, whether this extension works or not is not obvious. 

     As a first step, in this section, we systematically analyze 
empirical data for antiproton elastic scattering and 
total reaction cross sections off stable nuclei at the incident 
energies of antiprotons of lower than about 1000 
MeV ~\cite{okinawa,apr09}.  We here focus on the case of 
antiprotons on C, because only in this case the empirical 
values of both the differential cross sections of elastic scattering 
and of $\sigma_R$  
are available.  As in the same way as the case of proton 
projectiles, the values of $\sigma_{\rm BS}$ are obtained from 
the first peak position of the elastic scattering data. 
Note that we regard the empirical values of absorption 
cross sections as those of $\sigma_R$. 

     As for empirical data for the elastic differential cross 
sections, Garreta {\it et al.}\ measured the angular distributions 
for elastic scattering of antiprotons from $^{12}$C at the incident 
kinetic energy, $T_{\bar p} = $ 46.8 MeV~\cite{garreta1}, and from 
$^{12}$C, $^{40}$Ca, and $^{208}$Pb at $T_{\bar p} = $ 179.7 
MeV ~\cite{garreta2}.  We analyze the data of $^{12}$C 
and obtain the values of $\sigma_{\rm BS}$, which are 
plotted in Fig.~\ref{fig-pbar}.  We remark in passing that 
in the case of proton projectiles, there are no first peaks
that appear in the measured elastic differential cross 
sections at the same incident energies.  This reflects how
strongly antiprotons are attenuated in the target nucleus 
compared with protons.

     Ashford {\it et al}.\ measured the antiproton 
differential cross sections on Al, Cu, and Pb for two incident 
momenta, 514 and 633 MeV/$c$, but the separation of the 
elastic from inelastic contributions is 
incomplete~\cite{ashford:prc30}.  Therefore, we do not adopt 
these data for the present analyses.  Incidentally, a 
similar type of measurements was performed by Nakamura 
{\it et al}.\ for the differential cross sections of elastic scattering 
of antiprotons on C, Al, and Cu at six beam momenta 
between 470 and 880 MeV/$c$~\cite{nakamurak}. 

      For completeness, we mention other empirical 
differential cross sections of antiprotons.  Bruge 
{\it et al}.\ comparatively studied 
the elastic scattering from $^{16}$O and $^{18}$O isotopes 
at 178.4 MeV~\cite{bruge1}, and 
from deuterium at 179.3 MeV~\cite{bruge2}.  
Lemaire {\it et al}.\ measured the inelastic contribution 
from $^{12}$C and $^{18}$O at 50 and 180 MeV~\cite{lemaire}. 

     For comparison with $\sigma_{\rm BS}$, we plot in 
Fig.~\ref{fig-pbar} the empirical values of $\sigma_R$,
which are taken from the absorption cross section data
\cite{aihara,nakamurak,abrams}. 
From this figure, we find that even in the reactions 
involving antiprotons, the values of $\sigma_{\rm BS}$ are 
consistent with the empirical values of $\sigma_R$ within the 
uncertainties although the values of $\sigma_R$ are rather 
scattered in the region of low incident energies. 
These results support the relevance of the BS picture for 
antiprotons, while the analyses of the data for other hadronic 
probes are in progress.

     We remark that at much higher energies, there exist
various data for $\sigma_R$; for example, Denisov 
{\it et al.}\ measured the absorption cross sections for pions, 
kaons, protons, and antiprotons on Li, Be, C, Al, Cu, Sn, Pb, 
and U in the 6 to 60 GeV/$c$ momentum range~\cite{denisov}, 
while Carrol {\it et al.}\ measured the absorption cross 
sections for pions, kaons, protons, and antiprotons on 
targets of Li, C, Al, Cu, Sn, and Pb at 60, 200, and 280 
GeV/$c$~\cite{carrol}.  These data could be of some use,
but are beyond the scope of the present work.

     In the energy region of interest here, the total 
antiproton-proton cross sections are about five times larger 
than $\sigma_{pp}^{\rm total}$, which implies a shorter 
mean-free path of an antiproton in the nuclear medium
than that of a proton at the same kinetic energy.  According 
to the results of ref.~\cite{BS3}, the BS radii for protons 
are located in the nuclear surface.  
Note also that the values of $\sigma_R$ in Fig.~\ref{fig-pbar} 
are four or five times larger than 
those of $\sigma_R(p+{^{\rm natu.}{\rm C}})$ 
as one can see from Figs.~\ref{edep-cpb} and \ref{figsqwell}.
Then, we can expect that 
the BS radii for antiprotons are located in a significantly 
outer surface region.  This will open up a possibility of 
studying the nuclear surface structure, which would 
control diffractive reactions in a different way for 
various hadronic probes.

\setcounter{equation}{0}
\section{Summary and Conclusion}
\label{summary}

In this article, 
we have found that a novel $A^{1/6}$ 
dependence plays a crucial role in systematically
describing the energy dependence 
of $\sigma_R(p+A)$ (Sec.~\ref{bs-adep}).
This finding, which is based on the BS cross-section
formula constructed from a simple optical depth
argument (Sec.~\ref{bs-adep}), exhibits a clear contrast with
the eikonal approximation with the square-well potential
(Appendix~\ref{sec-sqwell}).

     The BS approximation of nuclei can be 
straightforwardly extended to other hadronic probes such as 
neutrons, antiprotons, pions, and kaons.  We have shown that 
the case of antiprotons works well 
(Sec.~\ref{other-probes}).  
We can expect that 
an antiproton is sensitive to the outer surface of nuclei than 
a proton, because the antiproton-nucleon total cross section 
is relatively large~\cite{apr09}.  It is interesting to
note the possibility that various hadronic probes of various 
incident energies could have the corresponding BS
radii cover the whole surface region.

    We can also extend the present framework to the case
of nucleus-nucleus reactions~\cite{BS3,kio08}.  This is
essential for the analyses of experimental data of
neutron-rich unstable nuclei measured or to be measured at
radioactive ion beam facilities such as the RI Beam Factory
of RIKEN Nishina Center~\cite{ozawa:npa693,takechi}.
We expect that the neutron-excess dependence of
empirical $\sigma_R(A+A)$ would play an important
role in deducing the density dependence of nuclear symmetry
energy~\cite{oyaii}.  
For description of the reactions between heavy nuclei,
however, it would be essential to allow for the
contribution from the Coulomb dissociation, which
would require additional treatment beyond the BS
approximation.  Studies in such new directions are
now in progress.

Recently, Horiuchi {\it et al}. 
examined the sensitivity of $\sigma_R$ 
to the neutron-skin thickness 
for O, Ne, Mg, Si, S, Ca, and Ni isotopes 
including neutron-rich unstable nuclei 
by performing numerical ``experiments" 
that adopt the Glauber approximation 
with the density distributions obtained 
from 
the Skyrme-Hartree-Fock method.~\cite{hyi:prc89}
They discussed 
such a sensitivity in terms of 
the reaction radius, $a_R$ $= \sqrt{\sigma_R/\pi}$. 
For the case of reactions with protons, 
$a_R$ is essentially the same as the BS radius $a$ 
introduced in Sec.~\ref{bsm}.
They found 
expressions for $a_R$
that linearly relate 
$a_R$ to the point matter radius
and the skin thickness
with energy dependent coefficients. 
Their elaborate study will offer us a great insight 
when we consider 
extension of our study toward a
further neutron-rich regime.

\mbox{}\\

We acknowledge T. Nakatsukasa and T. Motobayashi for 
constructive comments and encouragement during the course 
of this work, M. Lantz and K. Yazaki for invaluable 
suggestions and comments, H. Kondo for helpful
cooperation and discussion, and the members of 
Hokkaido University Nuclear Reaction Data Centre (JCPRG)
(http://www.jcprg.org/), particularly N. Otsuka, 
for kindly helping us collect various data sets. 
Last but not least, we thank H. Iwase, S. Hashimoto, 
K. Niita, and other development members of PHITS 
for discussion on further application of our framework.
This work was supported in part by Grants-in-Aid 
for Scientific Research on Innovative Areas through
No.\ 24105008 provided by MEXT.

\appendix
\setcounter{section}{0}

\setcounter{equation}{0}
\section{Parametrizations of 
Proton-Nucleon Total Cross Sections} 
\label{sigmatotal}

     For the values of $\sigma_{pN}^{\rm total}$ in Eq.~(\ref{form0}), 
we adopt the parametrization proposed by Bertulani and De Conti, 
Eqs.~(1) and (2) of ref.~\cite{bertulani}.  For completeness, 
we simply summarize their expressions in the unit of mb as follows: 
\begin{equation}
   \sigma_{pp}^{\rm total} =
   \left\{
   \begin{array}
   [c]{c}%
   19.6 + {4253/T_p} - {375/\sqrt{T_p}} + 3.86\times10^{-2}T_p \\
   ({\rm for }\ T_p < 280\  {\rm MeV}) 
\\ \; \\
   32.7 - 5.52\times10^{-2}T_p + 3.53\times10^{-7}T_p^3  \\
   -  2.97\times10^{-10}T_p^4  \\
   ({\rm for }\ 280\ {\rm MeV} \le T_p < 840\ {\rm MeV}) 
\\ \; \\
   50.9 - 3.8\times 10^{-3}T_p + 2.78\times 10^{-7}T_p^2 \\
   + 1.92\times 10^{-15} T_p^4  \\
   ({\rm for}\ 840 \ {\rm MeV} \le T_p \le 5 \ {\rm GeV})
   \end{array}
   \right.
\label{signn1}
\end{equation}
for proton-proton collisions, and
\begin{equation}
   \sigma_{np}^{\rm total} =
   \left\{
   \begin{array}
   [c]{c}%
   89.4 - {2025/\sqrt{T_p}} + {19108/T_p} - {43535/T_p^2} \\
    ({\rm for }\ T_p < 300\ {\rm MeV}) 
\\ \; \\
   14.2 + {5436/T_p} + 3.72\times10^{-5}T_p^2  \\
   - 7.55\times10^{-9}T_p^3 \\
   ({\rm for }\ 300\ {\rm MeV} \le T_p < 700\ {\rm MeV}) 
\\ \; \\
   33.9 + 6.1\times 10^{-3}T_p - 1.55\times 10^{-6}T_p^2 \\
   + 1.3\times10^{-10}T_p^3 \\
   ({\rm for}\ 700 \ {\rm MeV} \le T_p \le 5 \ {\rm GeV}) 
   \end{array}
   \right.
\label{signn2}
\end{equation}
for proton-neutron collisions.  Here, $T_p$ is the kinetic 
energy of the projectiles in the laboratory frame in 
the unit of MeV.  These expressions, which are
constructed by $\chi^2$ fitting in such a way as to reproduce 
the energy dependence of the empirical 
$\sigma_{pN}^{\rm total}$, are valid up to 5 GeV 
(see Fig.~1 in ref.~\cite{bertulani} for details).

     For comparison, we estimate $\sigma_{pN}^{\rm total}$   
using the SAID program of the version "SP07"~\cite{sp07,said}. 
The SAID program gives several estimations of the observables 
based on the partial-wave analyses of the latest compilation of 
nucleon-nucleon scattering data. 

In Fig.~\ref{figsigpn} in Sec.~\ref{overestimate}, 
we compare the two parametrizations for 
$\sigma_{pN}^{\rm total}$ with the empirical values 
as a function of nucleon incident energies lower
than 1.5 GeV.  
The solid curves 
show the fitting by Bertulani and De Conti~\cite{bertulani}, 
and the dashed curves are obtained by the SAID program. 

     As shown in Fig.~\ref{figsigpn}, the 
parametrization given by Eqs.\ (\ref{signn1}) and (\ref{signn2}) 
very well reproduces the empirical values for $T_p$ up to 
$\simeq 1.5$ GeV, except for the proton-proton scattering for the 
energies lower than 300 MeV in which case its deviation 
from the SAID parametrization is appreciable due partly to 
uncertainties in the empirical data.  Both parametrizaitons give 
an almost indistinguishable prediction of the proton-nucleus total 
reaction cross sections via Eq.\ (\ref{form0}), but in this work 
we adopt a simpler one, namely, Eqs.\ (\ref{signn1}) and 
(\ref{signn2}).  We remark in passing that the values of the SAID 
program start to rapidly deviate from the data in the downward 
direction beyond 1.5 GeV.

     Another parametrization was proposed by Charagi and 
Gupta~\cite{charagi}.  This works well for proton incident energies 
lower than around 300 MeV in the laboratory frame.  
For $\sigma_{pp}^{\rm total}$, however, the parametrization
underestimates the experimental values for the energies higher than 
around 700 MeV up to 1000 MeV and significantly overestimates 
them for the energies higher than around 1000 MeV.  Also for 
$\sigma_{pn}^{\rm total}$, the agreement with the empirical 
values is not good for the energies higher than 400 MeV. 
Therefore, we do not adopt the parametrization for the 
present work.

\section{Scattering with a Square-Well Potential of Finite Strength}
\label{sec-sqwell}

     In this Appendix, we derive several expressions that characterize
the scattering with a square-well potential of finite strength using 
various expressions that appear 
in Sec.\ 2 of ref.~\cite{kurosupple}.
We apply the expressions derived here to discussion of the $A$ 
dependence of $\sigma_R(p+A)$ in the main text. 
For additional expressions for scattering amplitudes
and absorption cross sections that arise from the
above potential, 
see Sec.\ 5
of ref.~\cite{kurosupple}.

\subsection{Simple Case}

     Let us consider the case of a complex-valued potential of 
finite strength 
\begin{equation}
    V_{\rm opt}(r) = (V_0 - i W_0) \theta(\bar{a} - r),
\label{recpot}
\end{equation}
where $V_0$ and $W_0 (> 0)$ are real constants, and $\bar{a}$ is
given by $A = \rho_0(4\pi/3) \bar{a}^3$, which is generally different
from the BS radius $a$ 
as in Eq.~(\ref{form-a}) (see also Eq.~(2.9) 
of ref.~\cite{kurosupple}). 
Here, for
simplicity, we assume the same potential cutoff scale $\bar{a}$ for 
neutrons and protons.  
Through the phase-shift function, $\chi(b)$,
defined by Eq.~(2.2) of ref.~\cite{kurosupple}, 
we write $\sigma_{\rm abs}$,
Eq.~(2.8) of ref.~\cite{kurosupple},
as 
\begin{eqnarray}
  \sigma_{\rm abs} 
                   &=& 2\pi \int_0^\infty bdb \; 
                      \{1 - |\exp[i \chi(b)]|^2 \}
\nonumber\\
     &=& \pi \bar{a}^2 C(\alpha) ~~(< \pi \bar{a}^2), 
\label{g-sigabs4}
\end{eqnarray}
where
\begin{equation}
  C(\alpha) = 1 - \frac{2}{\alpha^2} 
                  [1 - (\alpha + 1)\exp(-\alpha)],
\label{g-sigabs4b}
\end{equation}
with
$\alpha = 4W_0 \bar{a} /v$ (see Eq.~(2.81) in ref.~\cite{slyv}).
In the limit of complete absorption ($\alpha\to\infty$), 
Eq.\ (\ref{g-sigabs4}) reduces to the correct form $\pi \bar{a}^2$.  
An extension to the case of the different potential cutoff scales
between protons and neutrons will be described 
in Sec.~5.2 of ref.~\cite{kurosupple}.

     If we apply the $t\rho$ approximation to the optical potential, 
we obtain
\begin{equation}
   W_0 = \frac{1}{2} {\bar \sigma}_{pN}^{\rm total} \rho_0 v, 
\end{equation}
which leads to 
\begin{equation}
   \alpha = 2{\bar \sigma}_{pN}^{\rm total} \rho_0 \bar{a},
\label{alpha}
\end{equation}
where ${\bar \sigma}_{pN}^{\rm total}$ is given by  Eq.\ 
(\ref{sigtot-ave}).

\subsection{$A$ Dependence}
\label{eiko-adep}

     Here we examine the target mass-number dependence of 
the expression for $\sigma_{\rm abs}$ given by
Eq.~(\ref{g-sigabs4}).  Since 
\begin{equation}
   \bar{a} = \epsilon A^{1/3} \propto A^{1/3}, 
\end{equation}
with $\epsilon = [3/(4\pi\rho_0)]^{1/3}$, and
\begin{equation}
   \alpha = \gamma {\bar \sigma}_{pN}^{\rm total} A^{1/3} 
          \propto A^{1/3}, 
\end{equation}
with $\gamma = 2 \epsilon \rho_0$, we obtain 
\begin{eqnarray}
  \sigma_{\rm abs} &=& \pi \bar{a}^2 C(\alpha)
\nonumber\\
                   &=& \pi \epsilon^2 A^{2/3} 
               C\left(\gamma {\bar \sigma}_{pN}^{\rm total} A^{1/3}\right)
\nonumber\\
\nonumber\\
                   &=& \pi \epsilon^2 A^{2/3}
                    + \frac{2\pi\epsilon^2}
                           {\gamma {\bar \sigma}_{pN}^{\rm total}} A^{1/3}
          \exp\left(-\gamma {\bar \sigma}_{pN}^{\rm total} A^{1/3}\right)
\nonumber\\
                   & & -\frac{2\pi\epsilon^2}
                            {\gamma^2 ({\bar \sigma}_{pN}^{\rm total})^2} 
                       \left[1 - 
         \exp\left(-\gamma {\bar \sigma}_{pN}^{\rm total} A^{1/3}\right)
                       \right].
\nonumber\\
\label{g-sigabsadep}
\end{eqnarray}
Each term is ordered in powers of $A^{1/3}$ except 
exponential factors.  The term proportional to $A^{2/3}$ is 
independent of energy in contrast to Carlson's formula~\cite{carlson}, 
which will be briefly summarized in 
Sec.~6 of ref.~\cite{kurosupple}.

\end{document}